\documentclass{llncs}

\PassOptionsToPackage{hyphens}{url}

\usepackage{bbding}
\usepackage{cite}
\usepackage{amsmath,amssymb,amsfonts}
\usepackage{orcidlink}
\usepackage{graphicx}
\usepackage{textcomp}
\usepackage{xcolor}
\usepackage{todonotes}

\usepackage[hyphens]{url}
\usepackage{hyperref}
\usepackage[hyphenbreaks]{breakurl}
\usepackage{booktabs}
\usepackage{multirow}
\usepackage{subfigure}
\usepackage{amssymb}
\usepackage[mathlines]{lineno}
\usepackage{caption}
\usepackage[linesnumbered, ruled]{algorithm2e}

\usepackage{lipsum}
\usepackage{multirow}  
\usepackage{array}     

\usepackage{geometry}
\begin{document}

\title{SysVCoder: An LLM-Driven Framework for Systematic Generation of System-Level Design}

\author{
    Jian Zuo\inst{1} \and 
    Junzhe Liu\inst{1} \and 
    Xianyong Wang\inst{1} \and 
    Chen Liang\inst{1} \and 
    Navya Goli\inst{2} \and 
    Umamaheswara Rao Tida\inst{2}\orcidlink{0000-0002-9724-1585} \and 
    Zhenge Jia\inst{1}\orcidlink{0000-0002-0554-3608}\thanks{Corresponding author: zhengejia@sdu.edu.cn}\textsuperscript{(\Envelope)} \and 
    Zhaoyan Shen\inst{1}\orcidlink{0000-0001-9526-6634}\and 
    Mengying Zhao\inst{1}\orcidlink{0000-0001-7891-5436}  
}

\institute{
    School of Computer Science and Technology, Shandong University, Qingdao, China\\
    \email{\{202320822, junzheliu, wangxianyong, chenliang123\}@mail.sdu.edu.cn} \\ 
        \email{\{zhengejia, shenzhaoyan, zhaomengying\}@sdu.edu.cn}
    \and
    Department of Electrical and Computer Engineering, North Dakota State University, Fargo, USA\\
    \email{\{navya.goli, umamaheswara.tida\}@ndsu.edu}
}

\maketitle

\begin{abstract}

Recent advances in large language models (LLMs) have demonstrated strong potential in generating hardware designs using hardware description languages (HDLs) such as Verilog. 
However, existing LLM-based frameworks struggle to accurately capture the complexity of real-world architectural designs, particularly for large-scale systems with hierarchical, multi-level module instantiations. 
To address this issue, we present SysVCoder, an LLM-driven framework that enhances both the generation quality and efficiency of system-level design in Verilog. 
SysVCoder introduces a two-stage generation pipeline that leverages an intermediate representation to enable a more structured and accurate translation from natural language specifications to complex multi-module designs.
Furthermore, we incorporate a rule-based alignment mechanism and a domain-specific retrieval-augmented generation strategy (DS-RAG) to enhance functional correctness by grounding LLM outputs in domain knowledge.
We also present SysVDB, a comprehensive dataset comprising 60 system-level hardware designs along with their corresponding verification testbenches. 
Experimental results demonstrate that SysVCoder outperforms state-of-the-art frameworks such as CodeV and VeriGen by 30.7\% and 38.3\% in terms of functional correctness under the same base LLM.
Notably, SysVCoder achieves performance comparable to NVIDIA’s GPT-4-based VerilogCoder while using only a 7B-parameter model, reducing token consumption by 7.6× and synthesis latency by 37.5×.
Both SysVCoder and SysVDB are made public at {https://gitee.com/sdu-aes-lab/sysvcoder/}.


\end{abstract}

\section{Introduction}


Large language models (LLMs) have shown strong potential in accelerating agile hardware design tasks.
Among these tasks, one of the most promising directions is the automatic synthesis of hardware designs from natural language instructions~\cite{liu2024rtlcoder, gao2024autovcoder}.   
LLMs are used to directly translate functional descriptions written in natural language into hardware description language (HDL), such as Verilog or VHDL~\cite{zhao2024codev, lu2024rtllm}. 
%
%
%
Despite the remarkable progress of LLMs in generating high-level programming languages~\cite{liu2024exploring, yang2024exploring}, 
HDL imposes far more stringent requirements on logical structure~\cite{gao2024autovcoder}, making it difficult for LLMs to effectively learn and reason about the hardware description of the circuit design. 
Moreover, the availability of high-quality HDL datasets is considerably more limited than high-level languages, and the lack of standardized benchmark implementations further restricts the ability to evaluate LLM-driven HDL generation frameworks~\cite{zhao2024codev, gao2024autovcoder}. 

To address these issues, datasets and benchmarks such as VerilogEval~\cite{liu2023verilogeval}, 
RTLLM~\cite{lu2024rtllm}, and 
MG-Verilog~\cite{zhang2024mg} are introduced to address the scarcity of training data and enrich the diversity of available Verilog code. 
Building on these datasets, existing explorations to improve LLM performance in HDL code generation can be categorized into three perspectives: \textit{prompt engineering (PE), domain-specific fine-tuning (DSFT)}, and \textit{agent-based generation (AG)}. 
Prompt engineering aims to enhance code generation quality by modifying the phrasing and structure of prompts without altering the model’s parameters~\cite{chang2023chipgpt, lu2024rtllm, thakur2023autochip}. 
Domain-specific fine-tuning takes a more direct approach by continuing model training on Verilog datasets, with strategies ranging from full model fine-tuning~\cite{thakur2023benchmarking, dehaerne2023deep} to domain-adaptive pre-training~\cite{liu2023chipnemo, zhao2024codev, pei2024betterv, liu2024rtlcoder}. 
Agent-based generation leverages multiple specialized AI agents (based on full-parameter LLMs such as GPT-4) to synthesize HDL code. 
Representative works such as VerilogCoder~\cite{ho2025verilogcoder} and MAGE~\cite{zhao2024mage} construct different AI agents specialized in task planning, code synthesis, and syntax checking to collaboratively generate syntactically and functionally correct Verilog code.

Despite efforts to enable effective LLM-based Verilog code generation, three key challenges impede the adoption of LLM in practical EDA design flow: 
1) Existing frameworks are typically learned and evaluated on benchmarks composed of isolated, single-module Verilog designs, which fail to capture the complexity of real-world, system-level design involving hundreds of lines of code and multi-level instantiations; 
2) Current generation workflows generally translate natural language descriptions directly into Verilog code, 
reducing the overall interpretability, adjustability, and transparency of the generation process; 
3) Extensive state-of-the-art (SOTA) frameworks rely heavily on powerful but proprietary LLMs like GPT-4, while open-source and affordable models tailored for practical system-level design generation frameworks still lag significantly in performance.

To address these challenges, we propose {SysVCoder}, an LLM-driven framework to enable accurate and systematic generation of system-level, multi-hierarchical hardware designs.
Specifically, we propose a two-stage generation pipeline that bridges the gap between simple natural language inputs and system-level design through a novel intermediate representation named General Intermediate Representation (GIR). 
Unlike abstract syntax tree (AST)-like formats, GIR distills designs into fundamental hardware design variables annotated with concise one-sentence comments on module functionality and streamlined logic structures. 
In the first stage of generation (i.e., translating natural language descriptions to GIR), we propose an automated augmentation mechanism that generates high-quality, diverse description-GIR training pairs, and a GIR-specific instruction tuning method to strengthen the model’s understanding of system-level design structures.
In the second stage of generation (i.e., translating GIR to system-level design), we introduce a rule-based alignment method that converts GIR into implementation details to enhance the LLM's understanding on structured intermediate representations. 
Additionally, we integrate a domain-specific retrieval-augmented generation (DS-RAG) module, leveraging the GIR’s embedded comments to retrieve relevant single-module implementations for better generation quality.
%
Experimental results demonstrate that SysVCoder outperforms existing SOTA frameworks VeriCoder~\cite{wei2025vericoder} and VeriGen~\cite{thakur2024verigen} by 31.6\% and 38.3\% in terms of functional correctness on system-level design benchmarks. 
Remarkably, it achieves comparable function correctness with a parameter count of only 7B to the SOTA agent-based generation framework VerilogCoder~\cite{ho2025verilogcoder} (with GPT-4 as base model), with 7.6$\times$ fewer tokens consumed and 37.5$\times$ less generation latency. 
Our contributions are summarized as follows:
\begin{itemize}
    \item We propose SysVCoder, a LLM-driven framework for system-level, real-world hardware design generation.
    \item We introduce a two-stage generation pipeline based on GIR, which improves the LLM's synthesis capability on system-level design and enhances the transparency of the synthesis process.
    \item Experimental results show that SysVCoder outperforms SOTA frameworks across various benchmarks in terms of functional correctness and practical performance.
    \item SysVCoder framework and SysVDB are made open-sourced, including the data generation pipeline, training datasets, testbenches, and fine-tuned model weights, to facilitate future research and reproducibility.
\end{itemize}

\section{Background and Motivation}
\label{III_motivation}

Current research efforts toward effective LLM-based HDL generation generally focus on two key areas: high-quality dataset preparation and post-training techniques (such as fine-tuning and instruction tuning) to enhance the model’s understanding and generation capabilities.

\subsection{Datasets for Verilog Synthesis}
Several efforts have been made to construct high-quality Verilog design datasets along with corresponding testbenches to evaluate functional and syntactic correctness. 
For instance, the authors of~\cite{thakur2023benchmarking} collect Verilog code from open-source GitHub projects to create a large-scale dataset for LLM training. 
VerilogEval~\cite{liu2023verilogeval}, released by NVIDIA Research, offers a comprehensive evaluation suite comprising 156 problems sourced from HDLBits~\cite{HDLBits}, accompanied by a benchmarking framework for automated functional verification. 
RTLLM~\cite{lu2024rtllm} also contributes a benchmark of 29 RTL designs paired with natural language descriptions and their respective testbenches.

Despite these advances, most datasets and benchmarks focus on evaluating LLM performance on isolated, standalone Verilog modules. 
This narrow scope fails to reflect the complexity and scale of real-world hardware design scenarios, which often involve extensive Verilog codebases with multi-level instantiations and interdependent modules. 
These limitations underscore the need for more comprehensive and realistic benchmarks that assess LLMs in the context of large-scale, system-level Verilog design tasks.

\subsection{LLM-based Hardware Design Synthesis Framework}
Prior works in LLM-driven Verilog synthesis focus on improving generation quality can be summarized in three categories: 
%
1) Prompt Engineering (PE) improves the quality of LLM's synthesized code by adjusting the descriptions and structures of prompts.
This approach does not involve altering model parameters, making it efficient and low-cost. 
The attempts in PE include introducing a prompt template for more efficient hardware design~\cite{chang2023chipgpt}, developing a self-planning prompt engineering approach~\cite{lu2024rtllm}, embedding compiler error messages into prompts as multi-loop synthesis feedback~\cite{thakur2023autochip}, etc; 
2) Domain-Specific Fine-Tuning (DSFT) focuses on conducting further training on the pre-trained LLM using the Verilog dataset to obtain better performances. 
For instance, ~\cite{thakur2023benchmarking} uses full-parameter fine-tuning on LLMs using data collected from GitHub while ChipNeMo~\cite{liu2023chipnemo} developed by NVIDIA deploys a domain adaptive pre-training approach with their private data.
CodeV~\cite{zhao2024codev}, VeriThoughts~\cite{yubeaton2025verithoughts}, VeriCoder~\cite{wei2025vericoder}, RTLCoder~\cite{liu2024rtlcoder}, and AutoVCoder~\cite{gao2024autovcoder} improve LLM capability by invoking various parameter-efficient fine-tuning strategies; 
3) Agent-based generation leverages multiple specialized AI agents to synthesize HDL code~\cite{ho2025verilogcoder, zhao2024mage}. 
VerilogCoder~\cite{ho2025verilogcoder} proposed by NVIDIA constructs AI agents specialized in task planning, code synthesis, and syntax checking to collaboratively generate syntactically and functionally correct Verilog code.

\begin{figure*}[t]
\includegraphics[width = \textwidth]{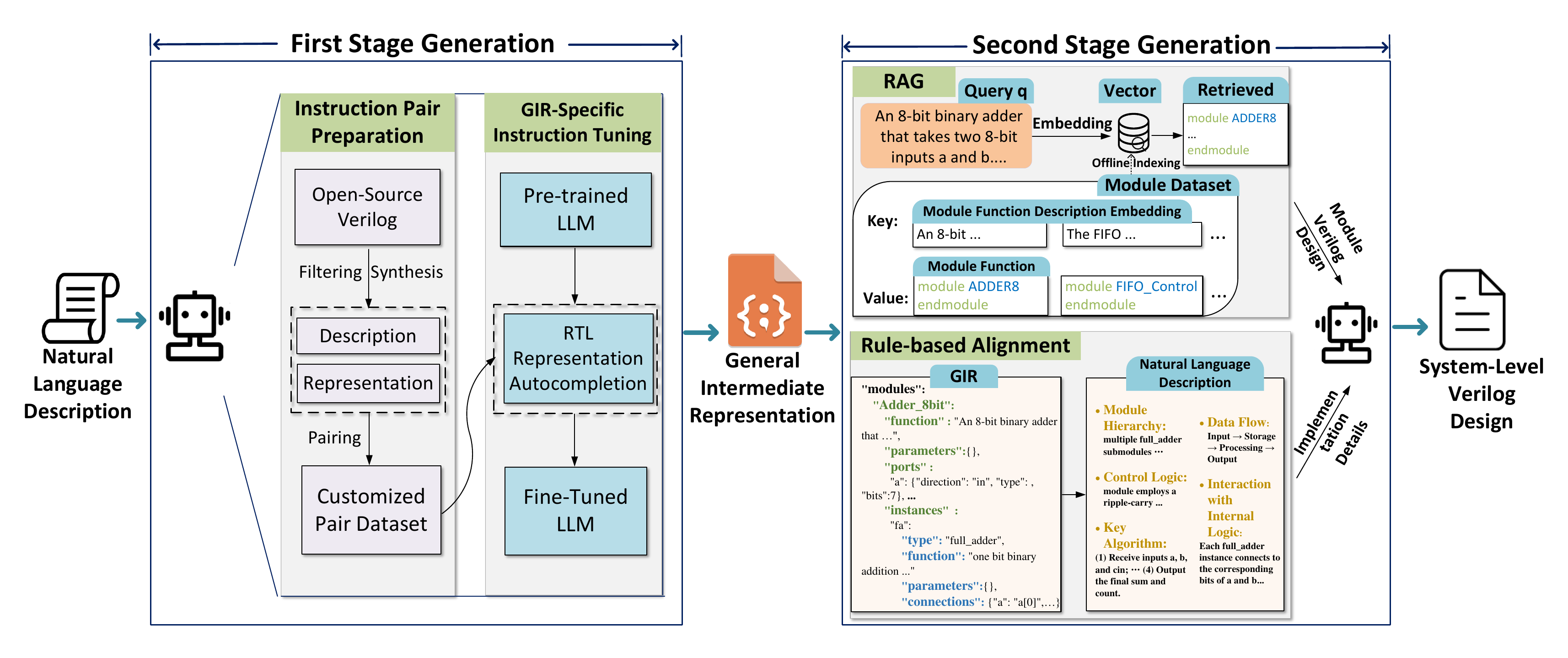}
\centering
\caption{
An overview of SysVCoder framework: (a) First stage generation from natural language description to intermediate representation, including the instruction pair preparation and GIR-specific instruction tuning approaches to enable the LLM to understand the natural language description and translate it into a structural representation; (b) Second stage generation from GIR to system-level hardware design, including retrain-free RAG and rule-based alignment approaches to enrich the input prompt to improve the LLM capability. 
}
\label{fig:overview}
\end{figure*}

Despite these advances, existing frameworks fall short when it comes to generating real-world, system-level Verilog designs, which often span hundreds of lines of code and involve multi-level module instantiations. 
Moreover, when presented with simple natural language descriptions of functional requirements in practice, the existing framework (even with the full-parameter LLMs such as GPT-4) struggles to generate correct and complete Verilog code. 
This highlights the pressing need for a more capable LLM-based framework that can reliably generate functionally correct Verilog designs from concise natural language inputs.

\section{System-Level Hardware Design Generation Framework}
\label{IV_method}


\label{subsec:llm_finetuning}


The system overview of SysVCoder is shown in Fig.~\ref{fig:overview}. 
SysVCoder adopts a two-stage generation approach, 
which leverages the proposed General Intermediate Representation (GIR) to break down the synthesis process, making it more manageable and transparent.
The GIR and the two-stage synthesis approach will be illustrated in the following sections. 


\subsection{General Intermediate Representation}
\label{subsec:ir_definition}

The General Intermediate Representation (GIR) introduced in this study serves as a critical abstraction layer designed to automate hardware design by bridging the gap between natural language descriptions and the system-level Verilog design. 
In the existing Verilog generation frameworks, LLMs are generally expected to infer hardware constructs (i.e., ports, instances, and signal connections) from unstructured text~\cite{gao2024autovcoder, liu2024rtlcoder} or structured but verbose instructions~\cite{zhang2024mg, zhao2024codev}. 
It often leads to syntactic errors or functional inaccuracies due to a lack of explicit structural guidance. 
Tailored specifically for LLMs, the GIR transforms unstructured, rough descriptions of the demanded function into a standardized, semantically rich format that is both human-readable and machine-processable. 

\begin{figure}[t]
    \centering
    \includegraphics[width=\textwidth]{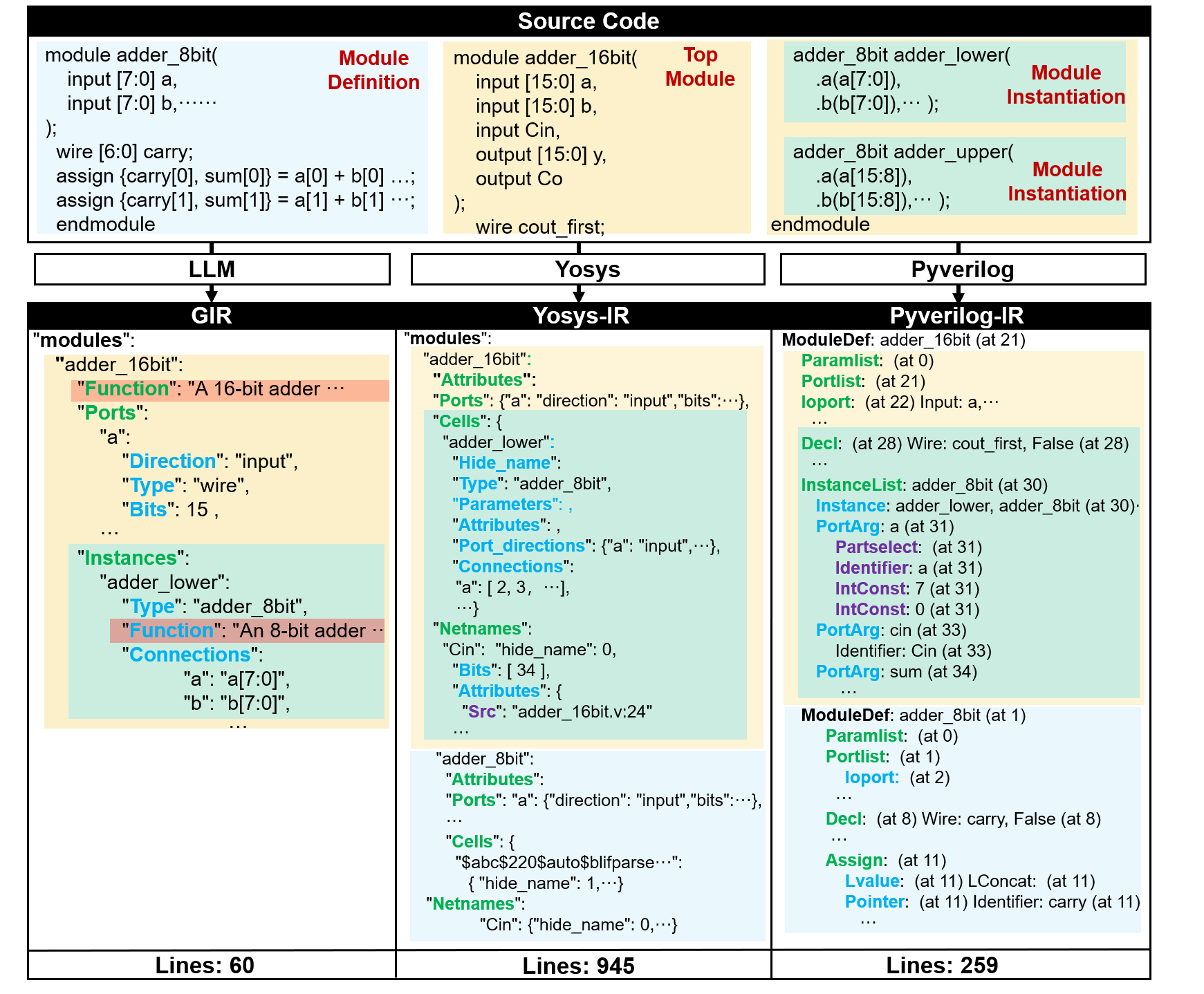} 
    \caption{Intermediate representation of \texttt{adder\_16bit} design by GIR, Yosys and Pyverilog. It turns out GIR achieves a more concise representation to describe the design. }
    \label{fig:ir_compare}
\end{figure}

As illustrated in Fig.~\ref{fig:ir_compare}, the GIR is structured in a hierarchical, JSON-like format that encapsulates the essential components of a hardware module, including fields such as \texttt{"modules"}, \texttt{"parameters"}, \texttt{"ports"}, \texttt{"instances"}, and \texttt{"connections"}. 
Each field is carefully defined to capture the architectural and functional semantics of hardware designs.
To ensure interoperability and rigorous analysis, the complete GIR schema is formally parameterized as a tuple $\mathcal{S}_{\text{GIR}} = \langle \mathcal{M}, \mathcal{R}, \mathcal{P}, \mathcal{I}, \mathcal{C} \rangle$. 
Specifically, \texttt{modules} ($\mathcal{M}$) acts as the root container defining the hierarchy and namespace scope for all local logic. 
Within this scope, \texttt{parameters} ($\mathcal{R}$) defines elaboration-time configuration constants for generic hardware templates, while \texttt{ports} ($\mathcal{P}$) delineates the module's I/O boundaries by explicitly specifying signal names, widths, and directions. 
For structural composition, \texttt{instances} ($\mathcal{I}$) tracks child module instantiations and their corresponding scope bindings. 
Finally, \texttt{connections} ($\mathcal{C}$) formalizes the netlist topology by mapping signal propagation from drivers to loads; this graph-based representation is critical for tracing dependency chains and identifying propagation paths of differential characteristics. 
Consequently, this normalized schema guarantees architectural fidelity while facilitating seamless automated parsing and algorithmic processing.

%
The conciseness of GIR is illustrated in Fig.~\ref{fig:ir_compare}, which compares its parsing of the \texttt{adder\_16bit} module with the intermediate representations (IR) generated via Yosys~\cite{wolf2013yosys} and Pyverilog~\cite{takamaeda2015pyverilog}. 
%
As shown in Fig.~\ref{fig:ir_compare}, existing IR designs remain closely tied to Verilog syntax.
To extract meaningful semantics from Pyverilog's AST, it requires significant post-processing since the output is difficult for LLMs to utilize directly.
It lacks semantic information and is heavily dependent on strict Verilog compliance, making it unsuitable for incomplete or iterative design workflows.
In contrast, GIR provides a concise, only 60-line representation that captures both the hierarchical structure and functional intent of the module. 
On the other hand, 
compared to a purely natural-language specification, 
the GIR introduces a highly interpretable and deterministic structure. 
While a natural-language description merely conveys functional intent, the GIR translates these implicit requirements into explicit structural boundaries. 
This semantic scaffolding allows LLMs to map text to hardware components without losing the flexibility of high-level abstractions.
It abstracts away low-level syntax in favor of modular, semantically meaningful fields that are easily populated and understood by LLMs. 
This structure enables generation through guided template completion rather than reverse engineering from verbose and syntax-bound outputs. 
Furthermore, the IR supports incremental refinement, allowing developers to gradually complete the design in a flexible and interpretable manner. 
Its JSON-like format aligns naturally with LLM tokenization strategies, facilitating seamless integration without the adaptation overhead.

\subsection{First Stage Generation}

In the first stage of generation shown in Fig.~\ref{fig:overview}, the goal is to enable the LLM to interpret a concise natural language description of a hardware design and translate it into GIR. 
To achieve this, we propose a GIR-specific instruction tuning method to teach the LLM domain-specific knowledge for accurately performing the translation.

We first collect 30k high-quality Verilog implementations from public datasets and open source projects. 
We then construct description-GIR pairs by leveraging Iverilog~\cite{williams2002iverilog} and commercial LLMs to generate natural language descriptions covering key module attributes such as name, function, parameters, and the intermediate representations that conform to the proposed GIR formatting rule. 
All GIR and the descriptions are manually rectified to avoid hallucination errors.
{\color{blue} 
(R1.1, 2.3)
}

Based on the prepared instruction pairs, we proposed a GIR-specific instruction tuning method to inject structured knowledge into LLMs. 
The method leverages the instruction samples to guide the model to generate detailed GIR representations from descriptive inputs.
The underlying learning mechanism of LLMs follows an auto-regressive paradigm. 
Let $\boldsymbol{z} = {z_1, z_2, \ldots, z_T}$ represent a training sequence comprising both an instruction and its response. 
An LLM parameterized by $\boldsymbol{\phi}$ estimates the probability of the entire sequence by factorizing it via the chain rule of probability~\cite{pei2024betterv}:

\begin{equation}
p_{\phi}(z) = \prod_{t=1}^T p_{\phi}(z_{t} | z_{j<t}).
\end{equation}
During fine-tuning, particularly with Low-Rank Adaptation (LoRA), the LLM is optimized to minimize the expected negative log-likelihood (or equivalently, cross-entropy loss) across a dataset $\mathcal{D} = {\boldsymbol{z}^{(1)}, \ldots, \boldsymbol{z}^{(N)}}$. The training objective is formalized as:

\begin{equation}
\mathcal{J}(\phi) = -\frac{1}{N} \frac{1}{T} \sum_{n=1}^{N}  \sum_{t=1}^T \log p_{\phi}(z^i_t | z^i_{j<t}).
\end{equation}
By integrating GIR as an intermediate representation of generation, we mitigate the synthesis difficulty imposed by the concise description against the complexity of  system-level Verilog design. 
The GIR serves as both a scaffold and a semantic guide, enabling more accurate and syntactically valid Verilog generation.

\begin{figure}[t]
    \centering
    \includegraphics[width=\textwidth]{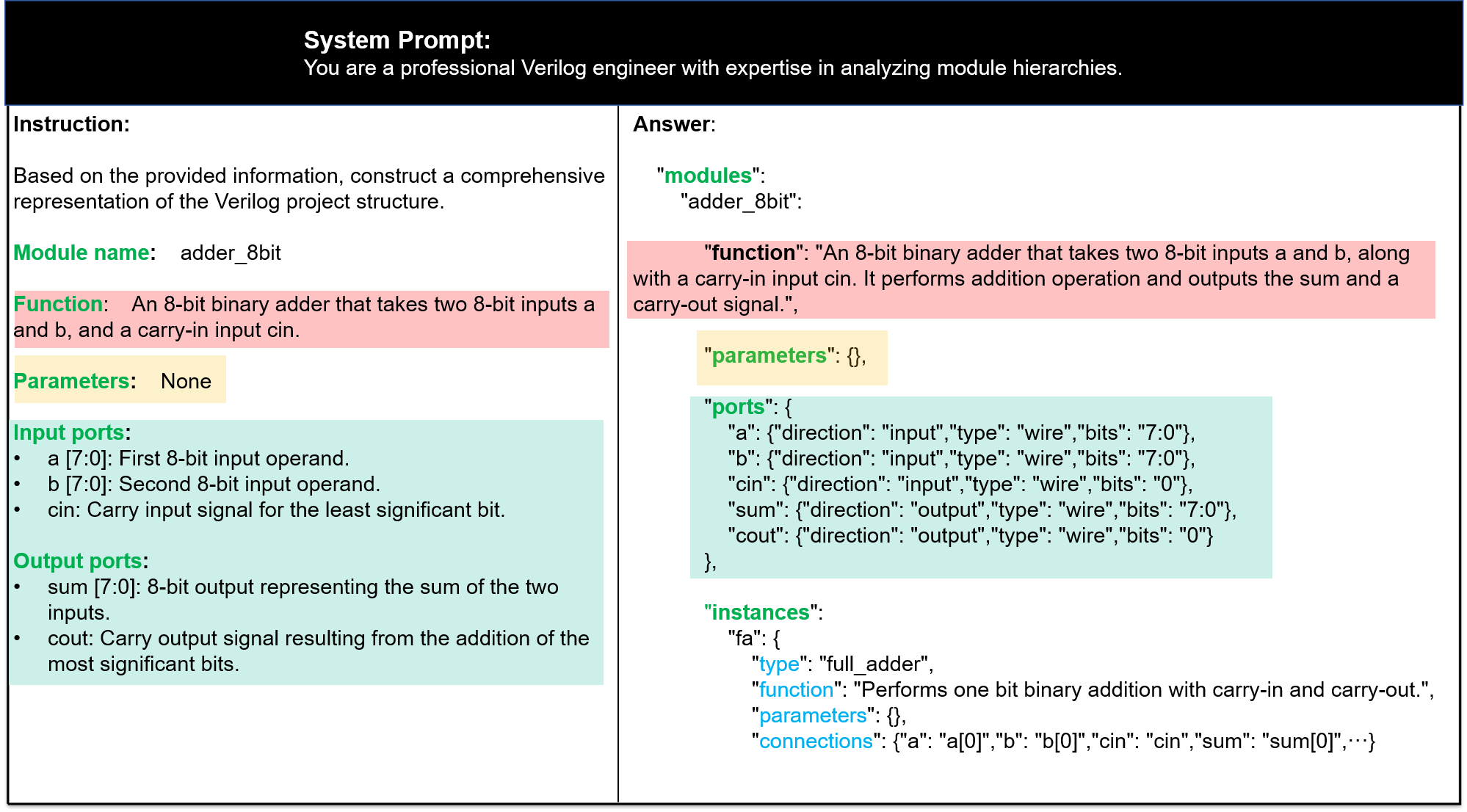}
    \caption{A case study to illustrate the instruction tuning process.  This example demonstrates how the LLM is guided to understand the natural language description of hardware functionality and generate its corresponding GIR.}
    \label{fig:instruct-tuning}
\end{figure}

\subsection{Second Stage Generation}


In the second stage of generation, as shown in Fig.~\ref{fig:overview}, the objective is to enable an LLM to generate system-level Verilog designs from GIR. 
Rather than relying on computationally expensive and time-consuming retraining on the retrieval model, we introduce a domain-specific Retrieval-Augmented Generation (DS-RAG) approach that combines rule-based alignment and a retraining-free retrieval to enhance prompt content to improve code generation performance.

\textbf{Rule-Based Alignment}. 
We first develop a rule-based natural language conversion mechanism that translates structured GIR data into rich, human-readable descriptions to enrich the quality of prompt input to the LLM. 
As shown in Fig.~\ref{fig:instruct-tuning}, the translated descriptions are utilized as part of the input prompt as detailed guidance to the LLM's generation process. 
This framework operates across three abstraction layers. 
At the top layer, the module definition captures the essential structure of the hardware design. 
The port description layer encodes individual port attributes. 
At the lowest layer, instantiation logic captures the hierarchical relationships and connectivity among modules.
To automate this conversion, we define a syntactic rule set that enables translation from GIR to structured natural language descriptions.
The core transformation rules are as follows:
\begin{itemize}
    \item \textbf{Module Declaration Rule}: Parses \texttt{ir\_data['modules']} into ``\texttt{<module\_name>} module defined with \texttt{<parameter\_count>} parameters, \texttt{<input\_count>} input ports, and \texttt{<output\_count>} output ports.''
    \item \textbf{Port Mapping Rule}: Generates for \texttt{ir\_data['modules'][ports]} the description ``Port \texttt{<n>} is of \texttt{<direction>} type, with bit-width \texttt{<width>}, synchronized to the \texttt{<clock\_domain>} clock domain.''
    \item \textbf{Instantiation Description Rule}: Converts \texttt{ir\_data['modules']['instances']} into ``Instantiates \texttt{<count>} submodules, where \texttt{<submodule\_name>} implements \texttt{<function\_description>}, connected to the parent module via \texttt{<interface\_name>}.''
\end{itemize}
The semantic descriptions generated through multi-stage translation are fed into an LLM to guide the production of implementation details covering the following aspects:
\begin{itemize}
    \item \textbf{Architecture Decomposition}: Include a module hierarchy matrix, a key submodule function mapping table, and storage configuration parameters.
    \item \textbf{Control Logic}: Define a state machine with \texttt{<state\_count>}, timing control based on <synchronous/asynchronous> reset mechanisms, and finite state machine encoding for processing steps.
    \item \textbf{Data Path}: Define a signal flow architecture from input buffers to pipeline registers, computational units, and output latches, incorporating an \texttt{<n>}-stage pipeline design with a bit-width conversion strategy named \texttt{<strategy\_name>}.
\end{itemize}

\textbf{Retraining-Free Retrieval}.
RAG retrieves relevant data from a database and integrates it into the prompt, enabling an LLM to generate accurate and contextually appropriate code.
Prior approaches such as ChipNeMo~\cite{liu2023chipnemo} and AutoVCoder~\cite{gao2024autovcoder} achieve this by retraining the retriever model to improve the quality of retrieval. 
However, preparing the required fine-tuning datasets and executing the retraining process is labor-intensive and computationally demanding.

To overcome this challenge, we propose a retraining-free retrieval mechanism that uses structured information in the GIR.
By leveraging natural language descriptions instead of Verilog code, we can effectively use a general embedding and retriever model without retraining the LLM.
As shown in Fig.~\ref{fig:overview}, we create a high-quality, single-module retrieval codebase from over 12,500 open-source Verilog and SystemVerilog projects. 
The code is revised using regular expressions to remove single-line comments, multi-line annotations, and conditional compilation directives. 
Independent modules with clear input/output interfaces are extracted, and their semantic function descriptions are converted into semantic vectors using a pre-trained embedding model. These are indexed with code fingerprints as \texttt{<code\_fingerprint, semantic\_vector>}.

During retrieval, the GIR-derived functional description is embedded using the same pre-trained model. 
A coarse-grained retrieval phase selects the top ten candidates by calculating cosine similarity between query and database vectors. 
A fine-grained re-ranking stage uses a pre-trained NLP Cross-Encoder~\cite{all-MiniLM-L6-v2} with the input format \texttt{[CLS] Query [SEP] Document [SEP]} to assess query-document pairs, improving relevance accuracy. 
Notably, the RAG mechanism retrieves contextually relevant individual modules to assist the generation, though discrepancies may exist between these retrieved source sub-modules and the actual target requirements. 
To reconcile these variations, the GIR leverages its structural \texttt{instance} metadata to enforce cross-module consistency, thereby guaranteeing the architectural correctness of multi-module connections and systemic interactions.
This RAG mechanism efficiently identifies contextually relevant code, boosting the LLM’s ability to generate high-quality Verilog code without retriever retraining.



\section{Experiments}
\label{V_experiment}


\begin{table}[t]
\centering
\setlength{\tabcolsep}{0.6mm}{
\caption{Metadata of SysVDB. }
\label{tab:benchmark_comparison}
\begin{tabular}{|l|c|c|c|}
\hline
\textbf{Benchmark} & \textbf{Dataset Size} & \textbf{Avg. Lines} & \textbf{Avg. Hierarchies} \\
\hline
VerilogEval-Human~\cite{liu2023verilogeval} & 156 & 22 & 1 \\
Problem-Set-VeriGen~\cite{thakur2024verigen} & 17 & 23 & 1 \\
RTLLM~\cite{lu2024rtllm} & 50 & 54 & 1.22 \\
SysVDB & 60 & 215 & 1.96 \\
\hline
\end{tabular}
}
\end{table}

\subsection{Experimental Setup}

\textbf{Dataset Preparation:} 
We construct an open-source dataset named \textit{SysVDB}, which is used to evaluate the capabilities of LLMs in generating system-level, real-world Verilog designs. 
As shown in Table~\ref{tab:benchmark_comparison} and Table~\ref{tab:module_classification}, 
SysVDB comprises 60 meticulously crafted Verilog designs, ranging from 7 different design domains.
To mitigate potential test-contamination concerns, we performe a de-duplication procedure between the 30K training corpus and the SysVDB benchmark. 
Specifically, all training samples are compared against SysVDB designs at three levels: (1) lexical similarity of functional descriptions; 
(2) structural similarity of the generated GIR representations;
(3) RTL-level similarity based on module hierarchy and behavioral signatures. 
Any sample exhibiting high similarity under any criterion is removed from the training set. 
Furthermore, GIR-description embedding analysis reveal that although several SysVDB tasks belong to commonly studied hardware categories (e.g., FIFO, UART, SPI, I2C, DMA, Cache, and FFT), their implementations, specifications, and module organizations differ substantially from the training samples. 
We emphasize that SysVDB is intentionally constructed from manually curated system-level designs emphasizing complex inter-module interactions rather than isolated IP blocks, thereby reducing benchmark memorization risk. 
Nevertheless, as with all evaluations based on open-source hardware designs, we cannot completely exclude the possibility that some concepts or implementation patterns appeared in the pretraining corpora of foundation models such as Qwen2.5 or DeepSeek. 
Importantly, all compared methods are built upon the same underlying foundation models and are evaluated on the same benchmark. 
Therefore, the substantial performance gains achieved by ComplexVCoder are attributable to the proposed GIR-based intermediate representation and two-stage generation framework rather than privileged access to benchmark-specific training data.

Each case in the SysVDB represents a realistic circuit design involving intricate inter-module interactions and structural hierarchies. 
In addition, each case is provided with a testbench and a comprehensive end-to-end pipeline to evaluate functional and syntactic correctness of the generated Verilog code. 
The evaluation employs line coverage as the primary metric. 
Beyond basic functional verification, the testbench incorporates corner cases, exceptional inputs, and signal integrity checks to ensure a robust and comprehensive assessment. 
A design is officially designated as "passed" only after successfully satisfying all verification checkpoints.

We further detail the cross-module interactions in SysVDB. 
Unlike existing benchmarks composed of isolated modules, SysVDB requires multi-module reasoning in three ways. 
First, it uses hierarchical parameter propagation, where a parent module overrides parameters to define the data widths and functions of its submodules during compilation. 
Second, it includes complex inter-module netlist topologies with multi-level dependency graphs; for example, signals in our security designs must pass through multiple hierarchy levels to connect with top-level control logic. 
Finally, submodules are connected via state-dependent protocols, such as multi-cycle handshakes, meaning an LLM must reason about joint timing behaviors across module boundaries rather than processing components individually. 
This structural density ensures that SysVDB functions as a system-level hardware benchmark.
To further assess the generalization ability of the SysVCoder framework, Problem-Set-VeriGen~\cite{thakur2024verigen} is included to further evaluate SysVCoder's generalization.

\begin{table}[t]
\renewcommand\arraystretch{1.2} 
\scriptsize                     
\centering
\setlength{\tabcolsep}{4pt}    

\caption{Module Types and Design Domain of SysVDB.}
\label{tab:module_classification}
\resizebox{0.98\textwidth}{!}{%
\begin{tabular}{|m{2.2cm}|m{2.8cm}|m{4.5cm}|} 
\hline
\textbf{Design Domain} & \textbf{Module Types} & \textbf{Verification Approach} \\
\hline
Bus \& Interconnect & AHB Master, APB Mux, DMA Controller & Verify AMBA protocol compliance, burst transfers, multi-master contention, error injection, and randomized corner cases. \\
\hline
Arithmetic Units & ALU, Multiplier, FFT & Check functional accuracy against golden models, boundary values, signed/unsigned operations, overflow, and formal equivalence. \\
\hline
Memory \& Storage & Cache, FIFO Memory, SDRAM Controller & Cover read/write consistency, cache hit/miss, clock domain timing, ECC, state machine transitions, and flag coverage. \\
\hline
Signal Processing & CIC Filter, Audio DAC, I2S Interface & Validate signal response, protocol alignment, sample rate variations, noise injection, and reference model comparison. \\
\hline
Communication & SPI Interface, UART, I2C Interface & Verify protocol transactions, error handling, arbitration, and edge cases with randomized packets. \\
\hline
CPU \& Control & Simple CPU, Interrupt Controller & Check instruction decoding, interrupt/exception handling, control flow, software co-simulation, and cycle-accurate timing. \\
\hline
Miscellaneous & Combination Lock, SIPO Shift Register, DAC & Validate state transitions, output conversions, functional paths, and behavior under valid/invalid inputs using directed and random tests. \\
\hline
\end{tabular}
}
\end{table}

\textbf{Baseline:} We evaluate the generation performances of SysVCoder against the following three categories of baselines: 
1) Commercial LLMs: we invoke general-purpose LLMs including \textit{GPT-4}, \textit{Deepseek-V3}, and \textit{GPT-3.5} to conduct Verilog generation; 
2) Domain-specific LLMs: we invoke SOTA domain-specific Verilog generation frameworks based on open-source LLMs, including \textit{codeV}~\cite{zhao2024codev}, \textit{RTLCoder}~\cite{liu2024rtlcoder}, \textit{ChipGPT}~\cite{chang2023chipgpt}, \textit{VeriGen}~\cite{thakur2024verigen}, 
\textit{MG-Verilog}~\cite{zhang2024mg}, \textit{VeriThoughts}~\cite{yubeaton2025verithoughts}, and \textit{VeriCoder}~\cite{wei2025vericoder};
3) Agent-based frameworks: we invoke multi-agent Verilog generation frameworks \textit{HiVeGen}~\cite{tang2025hivegen}, 
\textit{VerilogCoder}~\cite{ho2025verilogcoder} by NIVIDIA and \textit{MAGE}~\cite{zhao2024mage} for comparison. 
We adopt the widely used pass@$k$~\cite{chen2021evaluating} metric to evaluate Verilog code generation, which measures the probability that at least one out of k generated code samples successfully passes validation.
This metric can be calculated as:

\begin{equation}
\text{pass@}k = \mathbb{E} \left[ 1 - \frac{\binom{n-c}{k}}{\binom{n}{k}} \right], 
\end{equation}
where \(n\) is the total number of test attempts for the task, and \(c\) is the number of correct code generations for the task. 

\begin{table}[t]
\centering
\small
\renewcommand\arraystretch{1.1}
\caption{Generation performances of SysVCoder and SOTA frameworks over different benchmarks.}
\label{tab:benchmark_combined_results}
\setlength{\tabcolsep}{1mm}
\resizebox{\textwidth}{!}{
\begin{tabular}{@{}p{1.7cm}p{3.3cm}p{1.5cm}|cc|cc@{}} 
\toprule
\multirow{2}{*}{\textbf{Type}} & \multirow{2}{*}{\textbf{Model}} & \multirow{2}{*}{\textbf{Param}} & \multicolumn{2}{c}{\textbf{SysVDB}} & \multicolumn{2}{c}{\textbf{VeriGen}} \\ 
\cmidrule(lr){4-5}\cmidrule(lr){6-7}
& & & \textbf{pass@1} & \textbf{pass@5} & \textbf{pass@1} & \textbf{pass@5} \\ \midrule
\multirow{3}{*}{Commercial} & Deepseek-V3 & 671B & 33.3 & 51.6 & 76.5 & 88.2 \\
& GPT-3.5 & 175B & 26.6 & 45.0 & 58.8 & 64.7 \\
& GPT-4 & 1.8T & 35.0 & 51.6 & 76.5 & 88.2 \\ \midrule
\multirow{7}{*}{DS-LLMs} & CodeV~\cite{zhao2024codev} & 7B & 11.6 & 16.6 & 52.9 & 70.6 \\
& RTLCoder~\cite{liu2024rtlcoder} & 6.7B & 5.0 & 6.6 & 47.1 & 70.6 \\
& ChipGPT~\cite{chang2023chipgpt} & 7B & 8.3 & 11.6 & 47.1 & 64.7 \\ 
& VeriGen~\cite{thakur2024verigen} & 16B & 3.3 & 8.3 & 41.1 & 58.8 \\ 
& MG-Verilog~\cite{zhang2024mg} & 7B & 3.3 & 5.0 & 41.1 & 52.9 \\
& VeriCoder~\cite{wei2025vericoder} & 14B & 8.3 & 15.0 & 29.4 & 47.1 \\
& VeriThoughts~\cite{yubeaton2025verithoughts} & 7B & 10.0 & 18.3 & 58.8 & 70.6 \\
\midrule 
\multirow{3}{*}{Agent} 
& HiVeGen (GPT-4)~\cite{tang2025hivegen} & 1.8T & 36.6 & 48.3 & 88.2 & 94.1 \\
& VerilogCoder(GPT4)~\cite{ho2025verilogcoder} & 1.8T & 45.0 & 60.0 & 94.1 & 100.0 \\
& MAGE~(GPT-4)~\cite{zhao2024mage} & 1.8T & 46.6 & 63.3 & 100.0 & 100.0 \\ \midrule 
\multirow{4}{*}{Ours} & SysVCoder (Deepseek) & 671B & 43.3 & 61.6 & 94.1 & 100.0 \\
& SysVCoder (GPT-4) & 1.8T & 43.3 & 65.0 & 94.1 & 100.0 \\
& SysVCoder (Qwen2.5) & 14B & 40.0 & 63.3 & 64.7 & 76.4 \\
& SysVCoder (Qwen2.5) & 7B & 41.6 & 58.3 & 64.7 & 70.6 \\ \bottomrule
\end{tabular}
}
\end{table}



\textbf{Implementation Details:} 
We adopt different LLMs as the base models in SysVCoder, including commercial LLMs (i.e., Deepseek-V3, GPT-4), and open-source LLMs with smaller scale (i.e., Qwen2.5-7B, Qwen2.5-14B). 
We use Iverilog~\cite{williams2002iverilog} and Pyverilog~\cite{takamaeda2015pyverilog} to check the syntactic correctness of the generated Verilog solutions. 
If the design is syntactically correct, we proceed by running the corresponding test bench from the benchmark suite and comparing the output of the generated solution against the golden reference to verify its functional correctness.
During the instruction tuning process, we apply LoRA~\cite{hu2022lora} to enhance efficiency while preserving model performance. 
The learning rate \(\gamma\) is set to \(2 \times 10^{-4}\), and the models are trained for 3 epochs. Experiments are conducted using eight NVIDIA 4090 GPUs.
The implementations of SysVCoder and SysVDB are made public at {https://gitee.com/sdu-aes-lab/sysvcoder/}. 

\begin{figure}[t]
    \centering
    \includegraphics[width=0.85\textwidth]{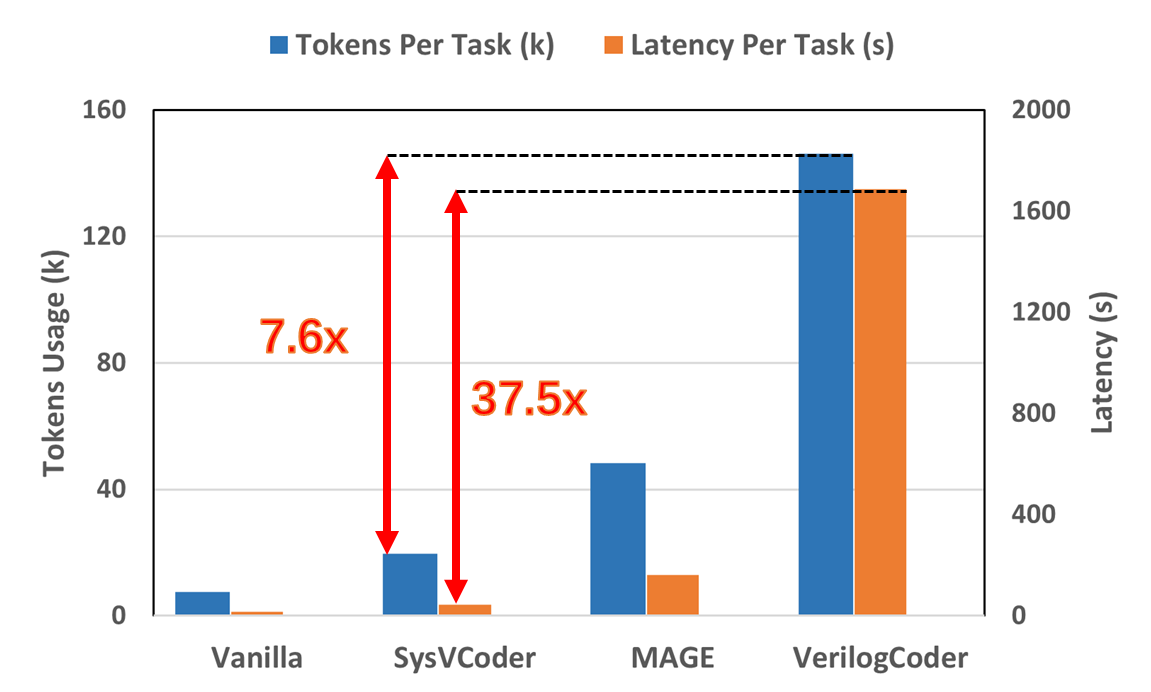}
    \caption{Comparison of average generation latency and tokens consumed by SysVCoder and agent-based approaches.}
    \label{fig:token_comparison}
\end{figure}

\subsection{Experimental Results}

We first evaluate SysVCoder's generation performance (in terms of pass@1 and pass@5) on system-level Verilog designs of SysVDB against baseline approaches. 
As shown in Table~\ref{tab:benchmark_combined_results}, 
SysVCoder substantially outperforms all existing generation frameworks with domain-specific LLMs. 
As shown in Table~\ref{tab:benchmark_combined_results}, SysVCoder with base LLM Qwen2.5-7B surpasses CodeV by over 30\% on pass@1 and 40\% on pass@5 with the LLM of a comparable scale. 
Moreover, SysVCoder with a 7B-LLM outperforms the VeriGen framework, which is based on a larger 16B-CodeGen LLM, by 38.3\% and 50.0\% in terms of pass@1 and pass@5 respectively. 
When compared with HiVeGen, the framework specifically designed for hierarchical Verilog generation based on GPT-4, SysVCoder with Qwen2.5-7B achieves 5\% and 10\% increase in terms of pass@1 and pass@5. 
Most existing Verilog generation frameworks conduct their training on isolated single-module Verilog examples, which limits their ability to handle system-level hardware designs in real-world scenarios.
These performances underscore SysVCoder's capability to effectively guide LLMs in understanding and generating complex circuit logic.

Notably, when compared with SOTA agent-based frameworks VerilogCover devised by NVIDIA, SysVCoder(QWen2.7-7B) achieves comparable performances, with only a 3.33\% drop in pass@1 and 1.65\% drop in pass@5, as shown in Table~\ref{tab:benchmark_combined_results}. 
In addition, as illustrated in Fig.~\ref{fig:token_comparison}, SysVCoder consumes 7.6$\times$ fewer tokens and generates 37.5$\times$ faster than VerilogCoder, which requires an iterative generation process with multiple full-scale LLM agents. 
Remarkably, SysVCoder delivers comparable performance through a streamlined two-stage generation process, effectively obviating the need for a syntax-failure retry loop.
This highlights that SysVCoder can significantly reduce the resource demand to enable lightweight LLMs to efficiently produce high-quality, system-level Verilog code, offering a scalable and practical solution for real-world hardware design automation.


To further assess the generalization capability of SysVCoder, we evaluate its performance of syntax and functional correctness on the public dataset Problem-Set-VeriGen~\cite{thakur2024verigen}. 
As shown in Table~\ref{tab:benchmark_combined_results}, 
SysVCoder(Qwen2.5-14B) achieves a pass@5 of 76.4\%, the highest among all frameworks based on small-scale LLMs, while SysVCoder(Deepseek) achieves perfect generation performances as the other baseline approaches based on GPT-4 or Deepseek-v3. 
These results demonstrate the strength of SysVCoder in enabling LLMs to more effectively capture the intrinsic characteristics of Verilog design, ultimately leading to higher-quality and more functionally accurate code generation.

\subsection{Ablation Study}



Table~\ref{tab:performance_comparison} presents the results of an ablation study evaluating the contribution of the key optimizations of SysVCoder.
These optimizations include general intermediate representation (GIR), instruction tuning for GIR generation (IT), 
rule-based alignment (RBA), and retraining-free retrieval-augmented generation (RAG).
The study is conducted across SysVDB benchmark dataset. 
For comparison, we also include a vanilla baseline, where the LLM directly generates Verilog code without any enhancements.

As shown in Table~\ref{tab:performance_comparison}, 
SysVCoder consistently delivers the best performance across all base LLMs, indicating the effectiveness of the proposed optimizations.
The performance drop of \textit{SysVCoder w/o GIR} indicates the important role of the intermediate representation as the bridge to enable an accurate translation from concise natural language description to system-level hardware design. 
The performance gap between SysVCoder and \textit{SysVCoder w/o IT} indicates the critical role of structural guidance through instruction tuning.
Notably, the zero-shot Qwen-7B baseline employing only GIR scaffolding as a prompt template fails to achieve significant performance gains, underscoring the limitations of relying solely on prompt engineering without instruction tuning.
In addition, removing the rule-based alignment (\textit{SysVCoder w/o RBA}) in the second-stage generation of SysVCoder results in significant performance drops. 
This highlights the importance of enriching prompts with structured, semantically aligned content.
Similarly, removing RAG (\textit{SysVCoder w/o RAG}) leads to moderate but consistent declines in performance, underscoring the benefit of including high-quality, standard Verilog module references in pormpt during generation.

\begin{table}[t]
\centering
\scriptsize
\renewcommand\arraystretch{1.1}
\caption{Generation performances under different scheme settings on SysVDB.}
\label{tab:performance_comparison}
\setlength{\tabcolsep}{0.8mm}
\resizebox{\textwidth}{!}{
\begin{tabular}{@{}p{2.5cm}cccccccc@{}}
\toprule
\multirow{2}{*}{Method} & \multicolumn{2}{c}{Qwen-7B} & \multicolumn{2}{c}{Qwen-14B} & \multicolumn{2}{c}{GPT-4} & \multicolumn{2}{c}{Deepseek} \\ \cmidrule(lr){2-3}\cmidrule(lr){4-5}\cmidrule(lr){6-7}\cmidrule(lr){8-9}
 & pass@1 & pass@5 & pass@1 & pass@5 & pass@1 & pass@5 & pass@1 & pass@5 \\ \midrule
Vanilla & 13.3 & 20.0 & 18.3 & 22.6 & 35.0 & 51.6 & 33.3 & 51.6 \\ 
SysVCoder & \textbf{41.6} & \textbf{58.3} & \textbf{40.0} & \textbf{63.3} & \textbf{41.6} & \textbf{60.0} & \textbf{43.3} & \textbf{61.6} \\ 
SysVC w/o GIR & 15.2 & 30.0 & 31.6 & 48.3 & 28.3 & 51.6 & 30.0 & 53.3 \\ 
SysVC w/o IT & 23.3 & 33.3 & 26.0 & 38.3 & - & - & - & - \\ 
SysVC w/o RBA & 28.3 & 45.0 & 31.6 & 55.0 & 35.0 & 53.3 & 38.3 & 50.0 \\ 
SysVC w/o RAG & 26.6 & 43.3 & 30.0 & 51.6 & 33.3 & 55.0 & 35.0 & 56.6 \\ \bottomrule
\end{tabular}
}
\end{table}

\section{Conclusion}
\label{VII_conclusion}

In this paper, we propose an LLM-driven framework, SysVCoder, for systematic generation of system-level hardware design. 
SysVCoder aims to generate real-world Verilog designs via the proposed two-stage generation mechanism, leveraging an intermediate representation to enable a more accurate and structured transition from natural language descriptions to intricate Verilog code. 
We also introduce SysVDB, a comprehensive dataset of 60 system-level Verilog designs with corresponding verification benchmarks derived from real-world implementations.
Experimental results show that SysVCoder outperforms SOTA Verilog generation frameworks and enables the small-scale LLM (i.e., Qwen-7B) to achieve comparable performances against agent-based frameworks with much less token consumption and generation latency.



\subsubsection*{Acknowledgments.} This study is supported in part by the National Natural Science Foundation of China (Grant No.62502279) and Young Talent of Lifting Engineering for Science and Technology in Shandong, China (Grant No.SDAST2024QTA078). 

\noindent 
The technical solution and experimental design presented in this paper were independently completed by the authors. AI tools were used solely for language polishing and formatting optimization, and did not participate in the development of the research ideas or core content.

\bibliographystyle{splncs04} 
\bibliography{./bib/ref.bib}

\end{document}